\documentclass[11pt,preprint]{aastex}
\usepackage{graphics} 
\usepackage{amssymb} 
\usepackage{amsmath} 
\usepackage[dvips]{color}

\newcommand{\A}{${\rm \AA~}$} 
 
\newcommand{\feii}{Fe\,{\footnotesize II}} 
\newcommand{\feiii}{Fe\,{\footnotesize III}} 
\newcommand{\mgii}{Mg\,{\footnotesize II}}

\newcommand{\civ}{C\,{\footnotesize IV}} 
\newcommand{\aliii}{Al\,{\footnotesize III}}

\newcommand{\siiv}{Si\,{\footnotesize IV}} 
\newcommand{\etal}{et al.~} 
\newcommand{\kms}{{km~s$^{-1}~$}} 

\begin{document}

\title{Strong Variability of Overlapping Iron Broad Absorption Lines in Five Radio-selected Quasars }

\author{Shaohua Zhang\altaffilmark{1}, Hongyan Zhou\altaffilmark{1,2}, 
Tinggui Wang\altaffilmark{2}, Huiyuan Wang\altaffilmark{2}, 
Xiheng Shi\altaffilmark{1,3}, Bo Liu\altaffilmark{2,1}, Wenjuan Liu\altaffilmark{2,1}, 
Zhenzhen Li\altaffilmark{2,1}, Shufen Wang\altaffilmark{2,1} }

\altaffiltext{1}{Polar Research Institute of China, 451 Jinqiao Road, Shanghai, 200136, China;
zhangshaohua@pric.org.cn; zhouhongyan@pric.org.cn}
\altaffiltext{2}{Key Laboratory for Research in Galaxies and Cosmology, Department of Astronomy, 
University of Science and Technology of China, Chinese Academy of Sciences, Hefei, Anhui, 230026, China}
\altaffiltext{3}{National Astronomical Observatories, Chinese Academy of Sciences, Beijing 100012, China}

\begin{abstract}
We present a study of the variability of broad absorption lines (BALs) in a uniformly radio-selected 
sample of 28 BAL quasars using the archival data from the FIRST Bright Quasar Survey (FBQS) and the Sloan 
Digital Sky Survey (SDSS) as well as data we obtained ourselves, covering time scales of $\sim 1-10$ years in the quasar rest frame. 
To our surprise, 5 quasars exhibiting strong variations all belong to a special subclass of 
`overlapping-trough' iron low-ionization BAL (OFeLoBAL) quasars; 
however, 4 other non-overlapping FeLoBALs (non-OFeLoBALs) are  invariant except in one case with a weak change in optical depth. 
Meanwhile, we also identify 6 typical variations of high-ionization and low-ionization BALs in this BAL quasar sample.
 Photoionization models suggest that OFeLoBALs are formed in  relatively dense ($n_e > 10^6$ cm$^{-3}$) outflows 
 at distances from the continuum source that range from the subparsec scale to dozens of parsecs. 
These formation conditions differ from those of non-OFeLoBALs,
which are likely produced by low-density gas located at distances of hundreds to thousands parsecs. 
Thus, OFeLoBALs and non-OFeLoBALs, i.e., FeLoBALs with/without strong BAL variations, may represent the bimodality of \ion{Fe}{2} absorption; 
the former are located in active galactic nucleus environments rather than in the host galaxy. 
We suggest that a high density and a small distance are the necessary conditions for the creation of OFeLoBALs. 
As suggested in the literature, strong BAL variability   is possibly attributable to variability in the covering factor 
of the BAL regions caused by clouds transiting across the line of sight   rather than  to variations in ionization. 
\end{abstract}
 
 
\section{Introduction}

Broad absorption lines (BALs; Weymann et al. 1991) are the most readily apparent signature of outflowing gas. 
  Strong outflows are believed to be one of the most important feedback processes connecting central supermassive black holes (SMBHs) to their host galaxies. 
However, the detailed physics of these outflows is not well understood. Studies of BALs can provide 
a useful   approach to placing constraints on the physical properties of outflows
  to further understand the overall picture of galaxy evolution and the connection 
between SMBHs and their surrounding hosts in quasars. 

BALs are observed in the blueward  of emission lines with a wide range of blueshifted velocities (up to $\sim  0.2c$) and broad widths (at least 2000 \kms) (Weymann et al. 1991). 
Based on the species of absorption, BALs are divided into three subclasses, i.e., high-ionization BALs (HiBALs), 
low-ionization BALs (LoBALs) and iron low-ionization BALs (FeLoBALs) (e.g., Hall et al. 2002). 
In addition to the BALs of the high-ionization species (e.g., \ion{C}{4}) in  HiBALs, and those of the low-ionization species (e.g., \ion{Mg}{2} and \ion{Al}{3}) in LoBALs,
FeLoBALs also show BALs in ground and excited of states \feii~and/or \feiii~(Hazard et al. 1987; Becker et al. 1997).
Statistical studies indicate   that approximately 10-20\% of quasars exhibit BALs, and $\sim$ 15\% of BAL quasars are detected as LoBAL quasars; 
  by contrast, FeLoBALs are the rarest subclass, appearing in at most $\sim$ 1\% of BAL quasars (e.g., Weymann et al. 1991; 
Reichard et al. 2003; Trump et al. 2006; Gibson et al. 2009; Zhang et al. 2010). 

BAL variability studies provide one method of assessing BAL structures, locations and dynamics, 
and   the results of such studies can potentially constrain the physical mechanisms responsible for outflows. 
In principle, the observed BAL variability could be explained by variations in the ionization parameter (e.g., Trevese et al. 2013), 
clouds transiting across the line of sight (e.g., Hall et al. 2011) 
or a combination of processes driven by these two factors. 
Recently, time variability in individual sources or samples with multi-epoch observations has been reported 
for HiBALs (e.g., Lundgren et al. 2007; Capellupo et al. 2012, 2013; Filiz Ak et al. 2013; Joshi et al. 2014), 
for LoBALs (Zhang et al. 2011; Vivek et al. 2014) and for FeLoBALs (Vivek et al. 2012; McGraw et al. 2014). 
Absorption-line variability is perhaps more common within shallower and higher-velocity troughs 
(e.g., Capellupo et al. 2011) on longer time scales (e.g., Gibson et al. 2008, 2010; Capellupo et al. 2011); 
occasionally,   absorption lines are even observed to disappear completely (e.g., Hall et al. 2011; Filiz Ak et al. 2012). 
However,   there has been no study to date that has compared variability among different BAL subclasses. 

Radio observations could   be used to constrain BAL orientations through investigations of radio variability, spectral index and morphology 
(Zhou et al. 2006; DiPompeo et al. 2012 and references therein). 
To date, only one BAL variability investigation of radio-loud (RL) BAL quasars   has been performed by Welling et al. (2014). 
They found that the BAL variability did not significantly depend   on either radio luminosity or radio-loudness; 
however, there was tentative evidence of greater fractional BAL variability within lobe-dominated RL quasars. 
The FIRST Bright Quasar Survey\footnote{The FBQS spectra are downloaded at ftp://cdsarc.u-strasbg.fr/pub/cats/J/ApJS/126/133.} 
(FBQS; White et al. 2000) provided a uniform radio-selected sample of 29 definite and tentative BAL quasars (Becker et al. 2000). 
  Using the data collected in this survey, it is possible to perform a statistical comparison of BAL variability among radio-selected BAL subclasses. 

This paper is organized as follows: \S 2 describes the multi-epoch observations of the sample, 
\S 3 presents the BAL variability, and \S 4 summarizes and discusses the implications. 
Throughout this paper, we adopt the CDM `concordance' cosmology with H0 = 70 km s$^{-1}$Mpc$^{-1}$, $\Omega_{\rm m} = 0.3$, and $\Omega_{\Lambda} = 0.7$.

\section{Sample Selection and Observations } 

White et al. (2000) presented the optical spectra of 636 quasars in the FBQS distributed over 2682 deg$^2$ 
in their Table 2 and Figure 16.   These quasar data form a heterogeneous set, collected at 5 different observatories between 1996 and 1998. 
The maximum coverage of the spectroscopic instruments is from $\sim$ 3600 to 10000 \A with a $\sim$4-10 \A resolution. 
We checked these spectra and adjusted the BAL quasar sample identified by White et al. (2000) and Becker et al. (2000). 
Two tentative BALs were rejected\footnote{FBQS J112220.5+312441 and J115023.6+281908 
do not meet the criterion of positive ``balnicity index'' (Weymann \etal1991).},
 one quasar (FBQS J105528.808+312411.65) was reclassified as a FeLoBAL quasar, and 4 LoBALs were reclassified as HiBALs and FeLoBALs. 
The final working sample contained 28 BALs, including 15 HiBALs, 4 LoBALs and 9 FeLoBALs (Table \ref{tab1}). 
  Of these, 5 FeLoBALs were discovered to exhibit abrupt dropes in flux, yielding almost no continuum windows below \ion{Mg}{2} 
and causing overlapping \ion{Fe}{2} troughs;   these were classified as `overlapping-trough' FeLoBALs (OFeLoBALs). 

We searched the repeated spectroscopic observations from the Sloan Digital Sky Survey (SDSS; York et al. 2000) databases 
  that have been obtained over the past twelve years, from May 2000 to July 2012. 
In the SDSS Seventh Data Release (DR7; Abazajian et al. 2009), 27 cases have spectra of higher spectral resolution. 
The DR7 spectrographs cover the wavelength range from 3800 to 9200 \A with a resolution of $\sim$ 1.5 \A at 6500 \A, 
which is nearly coincident with the wavelength coverage of the FBQS. 
  In addition, 16 cases were observed with the Baryon Oscillation Spectroscopic Survey (BOSS) spectrographs, 
whose spectral resolution varies from $\sim$ 1300 at 3600 \A to 2500 at 10,000 \A (Smee et al. 2013). 
These new repeated observations were released as part of the SDSS Tenth Data Release (DR10; Ahn et al. 2014). 

Meanwhile, we performed follow-up observations of FBQS J072831.6+402615 and J105528.8+312411 
using the Opto-Mechanics Research (OMR) spectrograph of the Xinglong 2.16 m telescope of 
the National Astronomical Observatories of China (NAOC) on March 24, 2009. 
A 200 \A mm$^{-1}$ dispersion grating and a PI $1340 \times 400$ CCD detector were used to 
cover a wavelength range of 3800-8500 \A at a 4 \A resolution. 
  Two exposures of 45 and 60 minutes each were acquired for both cases. 
A slit width of 2.5'' was chosen to match the typical seeing disk. 
For FBQS J072831.6+402615,   an additional spectroscopic observation was performed on Dec. 12, 2013, 
using the Yunnan Faint Object Spectrograph and Camera (YFOSC) instrument of 
the Lijiang 2.4 m telescope of the Yunnan Astronomical Observatories (YNAO). 
The blue-band grism, G14, provided a wavelength coverage from $\sim$ 3500 to 7500 \A with a spectral resolving power of 1337. 
The typical seeing was around 1.5'', the slit width was 2.0'', and one exposure of 30 minutes was acquired. 
The follow-up spectroscopic data were reduced using the IRAF longslit and echelle packages.

\section{BAL Variability }

The repeated spectra,   which were converted to the quasar rest frame using the redshifts reported in White et al. (2000),
cannot be directly compared with the FBQS spectra because of the absolute flux calibration or continuum variability. 
Thus, we used a power-law form ($\propto \lambda^{\alpha_{\lambda}}$) to scale the repeated spectra to match the FBQS spectra, 
  thereby masking the obvious absorption troughs.
However, we found that the power-law indexes were equal to zero in most cases. 
The large time intervals between the FBQS and repeated spectroscopic observations are comparable 
to the expected time scale for outflow rearrangement,   thus making the detection of variability more likely. 
We restricted our focus to BAL variability,   while neglecting both emission-line variability and residual-continuum variability. 

  Through spectral comparisons among two or three epochs, one can clearly identify variable absorption troughs in 12 BAL quasars, 
including 5 ``strong'' variations (Figure 1, 2 and 3) and 7 ``typical'' variations\footnote{BAL variations in FBQS J130425.5+421009 and J142013.0+253403 are tentative 
because the wavelength regions of variations are located at the blueward of their spectra or with low signal-to-noise ratio.} (Figure 4). 
In this work,   a typical variation refers to a small-amplitude velocity structure and/or optical-depth change of the absorption trough; 
by contrast, a strong variation is defined  as an extremely large and continuous change. 
In the figures, the FBQS spectra are plotted in black, and the repeated spectra are plotted in red and green. 
Table 1 suggests that BAL variability   does not tend to increase with greater radio-loudness or luminosity. 
This is consistent with the study of the BAL variability of RL quasars   performed by Welling et al. (2014),
  in which no evidence of any correlation of the radio-loudness and luminosity with the BAL variability was found; 
the BAL variations in RL BAL quasars appear to occur in a similar manner as those in radio-quiet BAL quasars. 
When we further explore the subtypes and structures of the BALs, 
  we find that the trends of BAL variability are completely different for HiBALs, LoBALs, FeLoBALs and OFeLoBALs. 
Typical BAL variations are detected in the first three subtypes; 
however,   intriguingly, all 5 instances of strong variations are detected in OFeLoBALs. 
Thus, we focus on strong variability in this work. 
In Figures 1, 2 and 3, we present the difference spectra for strong variations. 
The difference spectra are the observed spectra for the corresponding MJDs normalized to the highest fluxes among the multi-epoch observations; 
  these spectra reveal the absorption-line variability more clearly. Next, we analyze the properties of these cases. 

As shown in Figure 1, FBQS J072831.6+402615 and J140806.2+305448 (first reported in Hall et al. 2011) 
  exhibit the strong weakening and even disappearance of the \ion{Fe}{2} absorption lines. 
  We providentially observed the process of transformation from an OFeLoBAL quasar to a non-BAL quasar 
in the repeated spectroscopic observations of these quasars. 
For FBQS J072831.6+402615,   its spectrum from the FBQS epoch contains strong absorption lines from 
\ion{Mg}{2} and \ion{Fe}{2} multiplets, with a blueshifted velocity of $\sim$ 16,000 \kms. 
In the OMR spectrum acquired 6.85 rest-frame years later, the absorption troughs of 
\ion{Mg}{2} and \ion{Fe}{2} UV1 and UV2+3 had decreased in absorption depth by a factor of four. 
After an additional 2.88 rest-frame years, the residual overlapping troughs had completely disappeared in the YFOS spectrum. 
FBQS J140806.2+305448 was a spectacular OFeLoBAL quasar with a covering fraction of \ion{Fe}{2} absorption of $\sim$ 75 percent, 
but now, it is only a modestly absorbing LoBAL quasar. Hall et al. (2011) have found 
there was no significant absorption variability in early observations, one rest-frame year prior to the KECK/ESI observation in March 2000; 
however, the \ion{Mg}{2} trough outflowing at 12,000 \kms decreased by a factor of two, 
and \ion{Fe}{2} troughs at the same velocity disappeared over the next $\sim$ 7 rest-frame years. 
The SDSS DR10 spectrum shows a complete lack of \ion{Fe}{2} troughs (see also the HET spectrum in Hall et al. 2011). 

The top panels of Figure 2 show that the three spectra of FBQS J152350.4+391404 are similar to the intermediate epochs of FBQS J140806.2+305448, 
especially the KECK/ESI spectrum (MJD 53377; Hall et al. 2011). 
  Troughs of \ion{Fe}{2} multiplets are clearly visible in all three epochs. 
The difference spectra reveal strong variability in the ultraviolet \ion{Fe}{2} multiplets (\ion{Fe}{2} UV1 and UV2+3) 
and very slight variations in the optical \ion{Fe}{2} multiplets (\ion{Fe}{2} Opt8 and Opt6+7). 
Meanwhile, the red-band spectrographs of the SDSS and BOSS   captured a highly blueshifted velocity 
and an extremely broad absorption trough of H$\beta$   at approximately 4700 \A in the SDSS DR7 and DR10 spectra. 
Furthermore, a near-infrared spectrum   that is quasi-simultaneous with the SDSS DR10 spectrum, 
which was acquired by the TripleSpec spectrograph of the Hale 200-inch telescope at Palomar Observatory, 
suggests the existence of H$\alpha$ BALs. 
The bottom panels present the observed spectra of the H$\beta$ and H$\alpha$ regimes overlaid with the best-fit models 
and the normalized spectra of the blueshifted Balmer absorption troughs. 
  The fitting procedure has been described in detail by Dong et al. (2008);  therefore, we outline it only briefly here. 
The optical continuum from 3600 \A to 6900 \A is approximated by a single power law. 
\ion{Fe}{2} emission multiplets, both broad and narrow, 
are modeled using the I Zw 1 template provided by V{\'e}ron-Cetty et al. (2004). 
The emission lines are modeled as multiple Gaussians: two Gaussians for broad Balmer lines,
 and one Gaussian for [\ion{O}{3}]. 

As shown in Figure 3, the BALs of \ion{Fe}{2} multiplets are ambiguous or  severely overlapped in FBQS J105528.8+312411 and J140800.4+345125. 
The spectral variability in these two cases is deepening overall below \ion{Mg}{2} and blueward. 
  It is difficult to resolve the absorption multiplets based on their single observations, 
but it is possible to identify strong variations of the iron multiplets in their difference spectra. 
The triple-epoch spectra of FBQS J105528.8+312411 reveal that the \ion{Mg}{2} BALs are nearly saturated 
and exhibit no shift. The absorption structures are somewhat similar to those of FBQS J140806.2+305448 and J152350.4+391404. 
We present the identification of the ultraviolet and optical \ion{Fe}{2} multiplets in the difference spectra. 
For FBQS J140800.4+345125, we identify the absorption troughs of \ion{Mg}{2}, \ion{Al}{3} and \ion{Fe}{2} UV1 
(perhaps higher-velocity \ion{Mg}{2}) in the FBQS spectrum and mark them by blue diagonal line regions. 
However, the residual fluxes of the three troughs   are constant in all epochs. 
The normalized fluxes of these troughs in the difference spectra are $\sim$ 1. 
This broad absorption line system is invariant. 
  In addition, there are other broad-band absorption troughs in the difference spectra blueward of 
the invariant absorption troughs   noted above, 
which are surmised to represent the variable absorption of \ion{Fe}{2} UV62, UV1 and UV2+3 
with the same blueshifted velocity as the invariant absorption system; 
the iron absorption should be completely overlapping. Another possible explanation is 
that the \ion{Fe}{2} absorption may be undetected in the later spectra of FBQS J105528.8+312411 and J140800.4+345125. 
  The absorption variability may be an illusion caused by a variation in the strength of the \ion{Fe}{2} emission, 
or the apparent overlapping-trough absorption   may instead be attributable to reddening changes in the unusual reddening curves, 
  which may cause a rapid dropoff in the flux below a certain wavelength (e.g., Leighly et al. 2014). 

\section{Summary and Discussion }
In this work, we constructed a radio-selected BAL quasar sample from the FBQS, including 28 reclassified BALs, 
and   also considered subsequent spectroscopic observations of these quasars from the SDSS and two other instruments 
to explore the BAL variability among their dual-/triple-epoch spectroscopic observations. 
  Of the investigated sample, 27 BAL quasars   were observed in at least one epoch in the SDSS, 
and follow-up observations of 2 additional quasars were obtained. 
When we scaled the   subsequently acquired spectra to match the continua of the FBQS spectra using a power-law form, 
we detected variable absorption troughs in 12 BAL quasars through the spectral comparison, 
including 5 BAL quasars with strong variations in the BALs of \ion{Fe}{2} multiplets. 
  Intriguingly, all strong BAL variations were detected in OFeLoBALs; however, 
other BALs,   either with typical variations or without variations, were non-OFeLoBALs. 

  The apparently opposite trend was observed by Vivek et al. (2012, 2014); 
in that study, FeLoBALs were found to be less variable than HiBALs and LoBALs. 
Vivek et al. (2012) probed the time variability of 5 FeLoBALs spanning an interval of up to 10 years 
but only detected strong variations of \ion{Fe}{2} UV34 and UV48 in the spectra of SDSS J221511.93-004549.9. 
Based on photoionization models, de Kool et al. (2002b) showed that an \ion{Fe}{2}  column density 
  that is higher than that of \ion{Fe}{2} can be easily produced in a high-density outflow 
($n_e \geqslant 10^{10.5}$ cm$^{-3}$ for an ionization parameter of $U \simeq 10^{-2}$). 
However, the remaining 4 FeLoBALs did not exhibit any significant changes either in optical depth or in velocity structure. 
Two of them,   according to studies of their high-resolution spectra performed with VLT and KECK/HIRES, 
suggested an electron density of $10^{3.3\pm0.2}$ cm$^{-3}$ with a distance from the continuum emission source of 
$6 \pm 3$ kpc for SDSS J031856.62-060037.7 (Dunn et al. 2010) and $n_e \le 500$ cm$^{-3}$ with $D \sim$ 230 pc for 
SDSS J084044.41+363327.8 (de Kool et al. 2002b, also included in this work.). 

For the 5 OFeLoBALs with strong variations studied in this work, the physical properties of the outflow gas are different. 
In the rotating outflow model, the trough variability   arises from a structure in the BAL outflow moving out of our line of sight 
to the ultraviolet-continuum-emitting region of the quasar's accretion disk. 
Based on the size of that region and the time scale over which the absorption changes, 
Hall et al. (2011) constrained the BAL structure of FBQS J140806.2+305448 to a transverse velocity between 2600 and 22,000 \kms 
and a   line-of-sight velocity of 12,000 \kms; thus, the moving BAL structure was considered to be located approximately 1.7 to 14 pc 
from the black hole, nearly ten to ninety times farther than the H$\beta$ broad-line region. 
Meanwhile,   the application of the large-scale spectral synthesis code CLOUDY (c10.00; Ferland et al. 1998) 
in the extensive parameter space suggests that the absorbers of all OFeLoBALs considered in this work are ionized gases 
with a density of $ \ge 10^6$ cm$^{-3}$ and a column density of $\ge 10^{22}$ cm$^{-2}$. 
The absorbers in FBQS J140806.2+305448 and J152350.4+391404   are located at distances from the subparsec scale to dozens of parsecs. 
For FBQS J072831.6+402615, we can provide only an upper limit of $\sim$ 447 pc, with   considerable uncertainty. 
More details can be found in Appendix A. Similarly, McGraw et al. (2014) detected the BAL variability of low-ionization species 
(\ion{Fe}{2} and \ion{Mg}{2}) in 4 objects with a representative upper limit on the distance of the absorber from the center engines 
of $\sim$ 20 pc, with crossing speeds of $\gtrsim 500$ \kms. 

For the remaining FeLoBALs   considered in this work, previous works have reported different locations for the outflow gas. 
De Kool et al. (2002b) have suggested that the BAL absorbers in FBQS J084044.4+363328   consist of a high-velocity, 
low-density gas and that the scale of the distance from the black hole is two hundred parsecs. 
Analyses of the Keck/HIRES spectra of FBQS J104459.5+365605 and J121442.3+280329 have placed the outflow gas of the \ion{Fe}{2} absorbers at distances 
from the nucleus of 700 pc, with ne $\sim 4\times 10^4$ cm$^{-3}$, and 130 pc, with $n_e > 10^6$ cm$^{-3}$, respectively (de Kool et al. 2001, 2002a). 
For other FeLoBALs,   specifically non-overlapping FeLoBALs (non-OFeLoBALs), in the literature, the typical distances between the outflow gas 
and the nucleus  have been derived to be approximately 3 to 28 kpc (3C 191, Hamann et al. 2001; QSO J2359-1241, Korista et al. 2008; 
SDSS J0838+2955, Moe et al. 2009; SDSS J0318-0600, Dunn et al. 2010; collected in Table 4 of Lucy et al. 2014). 

Above all, FeLoBALs with strong variations are constrained to be   located in regions of high density ($\sim 10^6 - 10^9$ cm$^{-3}$) 
and close to the   continuum ionization sources (the upper limit is generally dozens of times farther than the size of the broad-emission-line region). 
However, the absorbers of other FeLoBALs   consist of low-density gas ($< 10^6$ cm$^{-3}$) and are located at distances of hundreds to thousands of parsecs 
from the central black holes. In other words, the former are located in active galactic nucleus (AGN) environments rather than in the host galaxy. 
Two subgroups, i.e., FeLoBALs with and without strong BAL variations, may represent the bimodality of \ion{Fe}{2} absorption observed by de Kool et al. (2002b). 
Our findings   regarding the different BAL variability behaviors of these two FeLoBAL subgroups suggest 
that they also correspond to overlapping and non-overlapping FeLoBALs. If our hypothesis is correct, 
then OFeLoBALs   correspond to densities that are ``higher by several orders of magnitude than in the absorbers in the `low-density gas at a large distance' group'', 
and ``their properties are best explained if they are 1000 times closer to the nucleus'' (de Kool et al. 2002b). 
On the other hand, we suggest that a high density of the absorbing gas and a large width of the absorption lines are the necessary conditions for OFeLoBALs (Hall et al. 2002). 
A higher density implies that   the absorbing gas should be located closer to the central engine, and higher densities and 
broader absorption lines certainly require smaller BALs;   these conditions are certainly consistent with the necessary conditions for variable BALs. 
The above estimations of the gas density and distance of strongly variable FeLoBALs are in good agreement with the requirements for OFeLoBALs.
   This is the reason why strong BAL variations are detected only in OFeLoBALs. 

The possible geometries and origins (Arav et al. 1994; de Kool \& Begelman 1995; Murray et al. 1995 and Elvis 2000)  imply that 
the variability of the BALs is explained by variations in the ionizing flux originating in the inner part of the accretion disk 
(e.g., Netzer et al. 2002; Trevese et al. 2013) and a transiting structure in the BAL outflow (e.g., Risaliti et al. 2002; Hall et al. 2011). 
The very optically thick gas   that is required for OFeLoBALs is very unlikely to   undergo any change in ionization that is sufficiently large 
for the gas to become only slightly optically thick and thus exhibit a detectable response to ionizing-continuum variability. 
Meanwhile,   the nine years of monitoring performed by the Catalina Surveys (Drake et al. 2014), 
beginning in April 2005, shows there is no apparent change in the optical continuum, at least   in the $V$ band, for these OFeLoBAL quasars (Table 1). 
  However, we cannot definitively rule out a change in the ionizing flux as the explanation for the strong variability of the \ion{Fe}{2} absorption. 
Observations   in the V band do not reflect the change in extreme UV and soft X-ray photons at $\lambda \leqslant 912$ \A; 
moreover, the ionizing and observed continua along our line of sight may not always track each other (Kaastra et al. 2014). 
If the BAL variability is related to the transverse motions of flows, then a large transverse velocity is required to produce 
such variability on a relatively short time scale, and the absorbing gas must be close to the black hole in a rotating disk wind model. 
This analysis agrees with Hall et al. (2011) that the strong BAL variability   is possibly attributable to variability 
in the covering factor of BAL regions caused by clouds transiting across the line of sight rather than to ionization variations. 
We also note that 3 of the 4 OFeLoBAL quasars   for which there are measurements of the radio spectral index are steep-spectrum radio sources, 
and the remaining one (FBQS J152350.4+391404) has a radio spectral index of $\alpha = - 0.4$, falling close to the dividing line, 
$-0.6 \leqslant \alpha \leqslant -0.4$ (Becker et al. 2000). The steep radio spectra suggest that 
  these quasars should be regarded as lobe-dominated radio sources   that are viewed at larger angles. 
This is also consistent with the interpretation of the strong BAL variations based on the rotating outflow model. 

\acknowledgments 
The authors are grateful to the anonymous referee for the helpful suggestions and the staff at the Lijiang 2.4 m telescope and the Xinglong 2.16 m telescope for the support during the observations. This work is supported by Chinese Natural Science Foundation (NSFC-11203021), National Basic Research Program of China (the ``973" Program, 2013CB834905) and the SOC program (CHINARE2014-02-03). TW acknowledge financial support from Chinese Natural Science Foundation (NSFC-11421303) and the Strategic Priority Research of Sciences (XDB09000000).

We acknowledge the use of the Lijiang 2.4 m telescope of the Yunnan Astronomical Observatories, the Xinglong 2.16 m telescope of the National Astronomical Observatories of China, and the Hale Telescope at Palomar Observatory through the Telescope Access Program (TAP), as well as the archive data from the FBQS, Catalina Surveys and SDSS.  Funding for the Lijiang 2.4 m telescope has been provided by Chinese Academy of Sciences and the People's Government of Yunnan Province.  The Xinglong 2.16 m telescope is supported by the open project program of the Key Laboratory of Optical Astronomy, National Astronomical Observatories, Chinese Academy of Sciences. TAP is funded by the Strategic Priority Research Program ``The Emergence of Cosmological Structures" (XDB09000000), National Astronomical Observatories, Chinese Academy of Sciences, and the Special Fund for Astronomy from the Ministry of Finance. Observations obtained with the Hale Telescope at Palomar Observatory were obtained as part of an agreement between the National Astronomical Observatories, Chinese Academy of Sciences, and the California Institute of Technology. Funding for SDSS-III has been provided by the Alfred P. Sloan Foundation, the Participating Institutions, the National Science Foundation, and the U.S. Department of Energy Office of Science. The SDSS-III web site is http://www.sdss3.org/.

\appendix

\section{ Physical Property Estimation  of the Iron Absorbers}
  The OFeLoBALs considered in this work are severely overlapping; thus, the geometry and physical conditions of the outflow winds 
cannot be accurately obtained   through analysis of the combination of the absorption lines.
Fortunately, we can   approximately evaluate the BALs by employing the large-scale spectral synthesis code CLOUDY (c10.00; Ferland et al. 1998) 
in the extensive parameter space. In the photoionization simulation, the geometry is assumed as a slab-shaped absorbing medium exposed 
to the ionizing continuum from the central engine with a uniform density, metallicity and abundance pattern. 
A typical active galactic nucleus multi-component continuum from Mathews \& Ferland (1987) is   specified as the incident ionizing radiation. 
The solar elemental abundance is adopted, and the gas is assumed to be free of dust.
   Each individual simulation model is  customized in terms of the ionization parameter ($U$), 
 electron density ($n_e$) and hydrogen column density ($N\rm_H$). 
The variation ranges in $U-n_e-N\rm_H$ space are $−3 \le {\rm log10}~U \le 0$, $4 \le {\rm log10} ~n_e~ ({\rm cm^{-3}}) \le 11$ 
and $20 \le {\rm log10}~ N\rm_H~ ({\rm cm^{-2}}) \le 23$, 
with a dex step of 1.0. The full 371-level Fe$^+$ model  implemented in CLOUDY,   which includes all levels up to 11.6 eV, 
is used to reproduce the \ion{Fe}{2} absorption. The employed model can predict the populations of various levels of Fe$^+$ and 
the strengths of the absorption lines arising from these levels. 

Figure A1 presents a series of models in $U-n_e-N\rm_H$ space. To clearly  illustrate the trend, one Gaussian profile, 
with its full width at half maximum ($FWHM$) set to 2000 \kms, is used to visualize the models. 
The calculations suggest that the iron multiplets raised from the ground state (e.g., \ion{Fe}{2} UV2+3 
  at approximately 2400 \A and \ion{Fe}{2} UV1 at approximately 2600 \A) are the most pervasive 
and that they are even generally saturated when the outflow winds have high column densities.
 However, the iron multiplets that are raised from the excited state   (e.g., \ion{Fe}{2} UV144-149 
and UV158-164 in the wavelength range of 2462-2530 \A   as well as \ion{Fe}{2} Opt6+7 and Opt8) are more sensitive 
to the   selected parameters. The iron absorption troughs increase with increasing Fe$^+$ density, 
as determined by the combination of the ionization parameter, the electron density and the hydrogen column density. 
  Moreover, for larger $U$ or $n_e$, more Fe$^+$ will become ionized to Fe$^{2+}$, and the absorption troughs of 
the high multiplets \ion{Fe}{2} UV144-149 and UV158-164   as well as \ion{Fe}{2} Opt6+7 and Opt8 will become even weaker.
 We can   approximately limit the physical parameters of the outflow winds based on the absorption strength/depth ratios 
of the \ion{Fe}{2} UV1 and UV2+3 multiplets, the \ion{Fe}{2} UV144-149 and UV158-164 multiplets, the \ion{Fe}{2} UV1 and UV2+3 multiplets, 
and the \ion{Fe}{2} Opt8 multiplets   as is done for the estimation of the strength ratios of \ion{Fe}{2} emission multiplets in Baldwin et al. (2004). 
The \ion{Fe}{2} UV144-149 and UV158-164 multiplets in four OFeLoBALs (with the exception of FBQS J105528.8+312411) 
  that are considered in this work have   depths that are nearly identical to those of the troughs of \ion{Fe}{2} UV1 and UV2+3, 
and in three cases, considerable absorption troughs of \ion{Fe}{2} Opt6+7 and Opt8   are evident. 
These characteristics suggest that the outflow winds of OFeLoBALs should be of high density ($n_e \ge 10^6~\rm cm^{-3}$) 
and high column density ($N\rm_H \ge 10^{22}~ cm^{-2}$). 

For individual sources with relatively isolated absorption multiplets, 
we  attempt to match the observed maximal difference spectra using photoionization models with appropriate broadening widths 
and blueshifted velocities. The covering factor is set to the average depths of \ion{Fe}{2} UV1 and UV2+3. 
In the search for optimal photoionization models, the reduced χ2 is evaluated using data covering absorption rest frames of 2300 to 3400 \A, 
which contain the overall absorption features of the \ion{Fe}{2} multiplets mentioned above. 
  Because the photoionization simulations are performed on a large grid, the physical parameters of the optimal models 
are determined through approximation, and a simulation using incremental grids will accurately evaluate the observations. 
In the right-hand panels of Figure 1 and the top-right panel of Figure 2, the best-matched photoionization models for the observed difference spectra 
are represented by blue curves. For FBQS J140806.2+305448, a model with $U = 10^{-1}$, $n\rm_e = 10^{11} ~cm^{-3}$, $N\rm_H = 10^{23} ~cm^{-2}$ 
and $FWHM = 2000$ \kms exhibits absorption features that are quite consistent with the observations, 
  with the exception of the absorption between the troughs of \ion{Mg}{2} and \ion{Fe}{2} Opt8. 
  The underestimation of the absorption at approximately 2850 \A is attributed to \ion{Cr}{2} $\lambda$2835, 2840, 2849, 2860, 2867 
and \ion{Mg}{1} $\lambda$2851 (Shi et al., submitted). 
Similarly, a model with $U = 10^{-2}$, $n\rm_e = 10^7 ~cm^{-3}$, $N\rm_H = 10^{23} ~cm^{-2}$ and $FWHM = 3000$ \kms is the optimal photoionization model for FBQS J152350.4+391404, 
and there are also obvious \ion{Cr}{2} absorption lines  present at approximately 2850 \A in the difference spectra. 
Using the ionization parameter definitions and the inferred   values of the ionization photon emission rate, 
ionization parameter and density, one can determine the distance of the outflow winds from the central ionization source, 
which is found to be $R\rm_{BAL} \sim$ 0.16 pc for FBQS J140806.2+305448 and 34 pc for FBQS J152350.4+391404. 
Based on the absorption depths of \ion{Fe}{2} UV1 and UV2+3 and possibly \ion{Fe}{2} UV62, {\rm we obtain a} model spectrum 
using $U = 10^{-3}$, $n\rm_e = 10^6 ~cm^{-3}$ and $N\rm_H = 10^{22}~ cm^{-2}$ to match the difference spectrum of FBQS J072831.6+402615; 
the distance of the outflow winds is  then found to be $\sim$ 447 pc. 
However, the difference spectra of FBQS J152350.4+391404 and J072831.6+402615 indicate stronger absorption of the \ion{Fe}{2} UV144-149 
and UV158-164 multiplets than   is present in the optimal models,
 which most likely suggests a higher density and a higher ionization parameter or 
some microturbulence in the outflow winds (Shi et al., submitted).
 If the density and/or ionization parameters of these two cases are underestimated, then the absorbing gas 
should be located closer to the ionization continuum sources than  our estimates indicate. 

\begin{deluxetable}{cc cc cc cc cc cc cl} 
\tabletypesize{\tiny} 
\rotate
\tablecaption{The properties of FBQS BAL quasars
\label{tab1} } 
\tablewidth{0pt} 
\startdata 
\hline 
Name (FBQS)&$z$&$M_{\rm B}$&log $L_{\rm R}$& log $R^*$&$\alpha$&MJD&MJD-plate-fiber&MJD-plate-fiber&$\Delta t^{1-2}$&$\Delta t^{1-3}$&$\Sigma\rm_V$&$\sigma\rm_{V}$&Type\\ 
(1) & (2) &(3) &(4) &(5) &(6) &(7) &(8) &(9) &(10) &(11) &(12) &(13) &(14)\\
\hline 
J072418.4+415914&1.552&-26.6&32.6&1.52&+0.0&51146&53312-1865-261&               &2.31&      & & &LoBAL$^\dagger$    \\ 
J072831.6+402615&0.656&-28.0&32.2&0.57&-1.1&50774&54915&56654&6.85&9.73  &0.09&0.06&OFeLoBAL$^\ddagger$    \\ 
J080901.3+275341&1.511&-27.7&31.9&0.40&... &50851&52618-0930-135&55858-4457-0512& 1.93& 5.46& & &HiBAL    \\ 
J084044.4+363328&1.230&-27.2&31.7&0.39&-0.2&50838&52320-0864-149&               & 1.82&     & & &FeLoBAL  \\ 
J091044.9+261253&2.920&-27.6&33.1&1.65&-0.5&51160&53415-2087-352&               & 1.58&     & & &HiBAL    \\ 
J091328.2+394444&1.580&-27.4&32.1&0.65&-0.6&50094&52707-0937-569&               & 2.77&     & & &HiBAL    \\ 
J093403.9+315331&2.419&-28.5&32.8&0.88&-0.2&50401&53386-1941-600&               & 2.39&     & & &HiBAL$^\dagger$    \\
J094602.2+274407&1.748&-27.9&32.4&0.74&$<$-1.5&50466&53385-1944-026&            & 2.91&     & & &HiBAL    \\ 
J095707.3+235625&1.995&-27.4&34.1&2.64&-0.6&51172&53737-2298-409&               & 2.35&     & & &HiBAL    \\ 
J103110.6+395322&1.082&-25.8&31.8&1.04&-0.2&50465&52998-1428-608&               & 3.33&     & & &LoBAL    \\ 
J104459.5+365605&0.701&-26.2&32.2&1.30&-0.5&50486&53463-2090-329&55615-4635-0704& 4.79& 8.26& & &FeLoBAL  \\ 
J105427.1+253600&2.400&-28.0&32.6&0.91&-0.5&50486&53793-2357-227&               & 2.66&     & & &LoBAL$^\dagger$    \\ 
J105528.8+312411&0.558&-23.9&31.7&1.68&... &50597&53711-2026-085&54915          &5.48&7.59&0.07&0.00&OFeLoBAL$^\ddagger$  \\ 
J120051.5+350831&1.700&-29.1&32.1&0.01&-0.8&50586&53469-2099-606&55648-4650-0052& 2.93& 5.14& & &HiBAL    \\ 
J121442.3+280329&0.698&-26.3&31.5&2.08&-0.8&50549&53823-2229-557&               & 5.28&     & & &FeLoBAL  \\ 
J130425.5+421009&1.916&-28.7&32.1&1.62&+0.7&50495&53119-1458-579&55681-4704-0881& 2.47& 4.87& & &HiBAL$^\dagger$    \\ 
J131213.5+231958&1.508&-27.4&33.3&1.30&-0.8&50585&54507-2651-380&56098-5977-0326& 4.28& 6.02& & &HiBAL    \\ 
J132422.5+245222&2.357&-27.4&32.8&1.32&-0.7&51009&54524-2664-537&56096-0186-0186& 2.87& 4.15& & &HiBAL    \\ 
J140800.4+345125&1.215&-26.3&32.0&1.01&-0.6&50486&53471-1839-377&55268-3855-0386& 3.69& 5.91&0.08&0.00&OFeLoBAL$^\ddagger$  \\ 
J140806.2+305448&0.842&-24.8&31.7&1.33&-0.7&50601&53795-2125-236&55276-3862-0124& 4.75& 6.95&0.07&0.00&OFeLoBAL$^\ddagger$  \\ 
J141334.4+421201&2.810&-28.3&33.5&1.69&-0.2&50586&52823-1347-539&56093-6059-0438& 1.61& 3.96& & &HiBAL    \\ 
J142013.0+253403&2.240&-27.9&32.1&0.49&-1.1&50596&53859-2127-011&56096-6015-0432& 2.76& 4.65& & &HiBAL$^\dagger$    \\ 
J142703.6+270940&1.170&-25.6&32.0&1.27&-0.7&50616&53876-2134-256&56067-6018-0412& 4.12& 6.88& & &FeLoBAL$^\dagger$  \\ 
J152314.4+375928&1.344&-26.5&31.9&0.78&-0.6&50245&53470-1400-450&56045-4979-0218& 3.77& 6.78& & &HiBAL    \\ 
J152350.4+391404&0.657&-26.3&31.6&0.61&-0.4&50507&52765-1293-234&56045-4979-0804& 3.73& 9.12&0.06&0.00&OFeLoBAL$^\ddagger$  \\ 
J160354.2+300208&2.026&-27.9&33.7&2.04&-0.6&50507&53496-1578-534&55707-5009-0569& 2.71& 4.71& & &HiBAL    \\ 
J164152.2+305851&2.000&-27.2&32.4&1.01&+0.5&51009&52781-1340-073&55832-5201-0914& 1.62& 4.40& & &LoBAL    \\ 
J165543.2+394519&1.747&-27.5&32.8&1.38&-0.2&51009&52079-0633-353&56098-6063-0438& 1.07& 5.08& & &HiBAL$^\dagger$   
\enddata 
\tablenotetext{Note.} {
(2-5): Redshift ($z$), absolute B magnitude ($M_{\rm B}$), radio luminosity (ergs cm$^{-2}$ s$^{-1}$ Hz$^{-1}$) at a rest-frame frequency of 5 GHz (log $L_{\rm R}$) and K-corrected 5 GHz radio to 2500 \A~ optical luminosity ratio (log $R^*$) from White et al. (2000).
(6): Spectral index between 3.6 and 20 cm ($F_{\nu} \propto \nu^{\alpha}$) from Becker et al. (2000).
(7-9):  Modified Julian Date of the FBQS and follow-up observations, MJD=JD-2400000. We also present their plate and fiber, if the follow-up spectra come from the SDSS.
(10-11): Rest-frame years of the second and third epoch since the FBQS epoch.
(12-13): Variance of observed magnitudes ($\Sigma_{\rm V}$) and the intrinsic amplitude ($\sigma_{\rm V}$) at $V$-band calculated by the formalism in Ai \etal(2010). 
$\dagger$:  Typical BAL variability. 
$\ddagger$: Strong BAL variability.} 
\end{deluxetable} 

\figurenum{1} 
\begin{figure*}[tbp] 
\epsscale{1.0}
\plotone{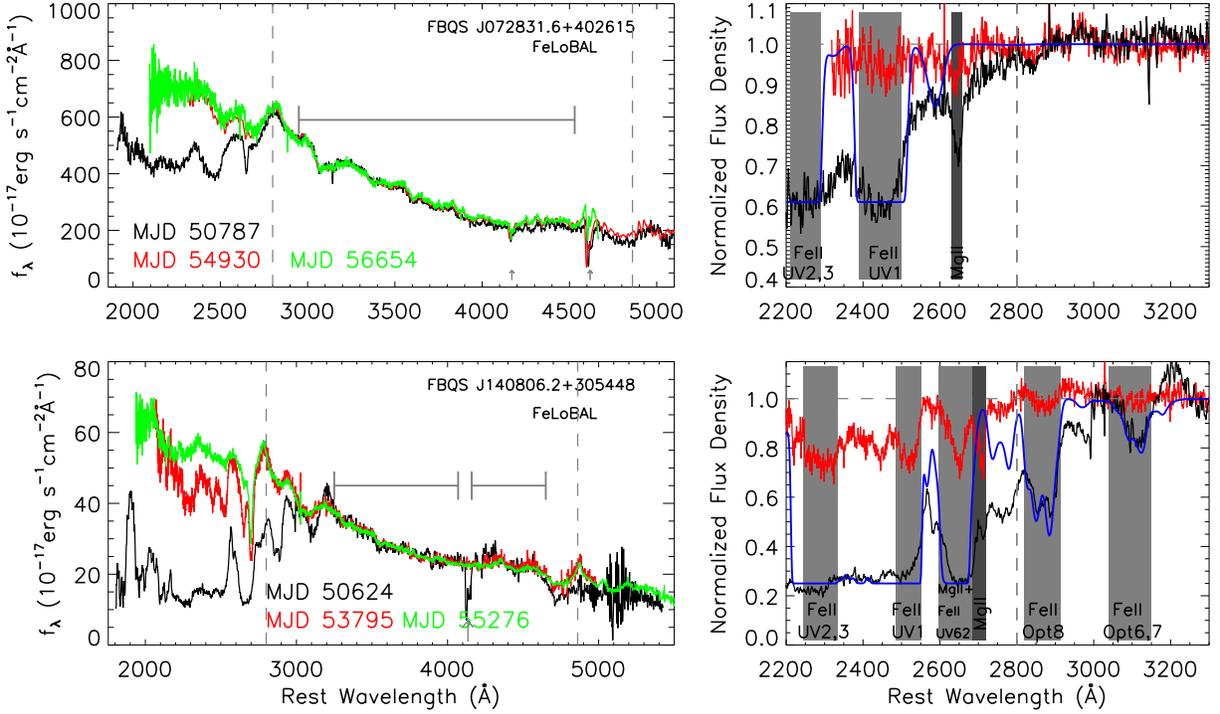}
\caption{Left: Multi-epoch observations of FBQS J072831.6+402615 and J140806.2+305448~in quasar's rest-frame. 
In each panel, flux scale applies to the repeated spectra to match the FBQS spectra in continuum 
and spectra have been smoothed with a three-pixel boxcar.
Solid horizontal lines show the wavelength coverages used in the spectrum scaled fitting.
Dashed vertical lines show the wavelengths of emission from H$\beta$ and \mgii.
Note that most of FBQS spectra have atmospheric absorption at $\sim$6880 and 7620 \A, masked by the arrows in the panels. 
Right: Difference spectra obtained from the corresponding spectra normalized by the highest fluxes of multi-epoch observations.
Dark gray shaded region shows the wavelength range of low-velocity \mgii~absorption.
Light gray shaded regions show the wavelength ranges absorbed by \feii~multiplets UV1 and UV2+3
and by a blend of \feii~UV62 and high-velocity \mgii~absorption.
Normalised model profiles for individual photoionization models broadened using Gaussian profile are also displayed by blue curves.
}\label{fig1} 
\end{figure*} 

\figurenum{2}
\begin{figure*}[tbp]
\epsscale{1.0}
\plotone{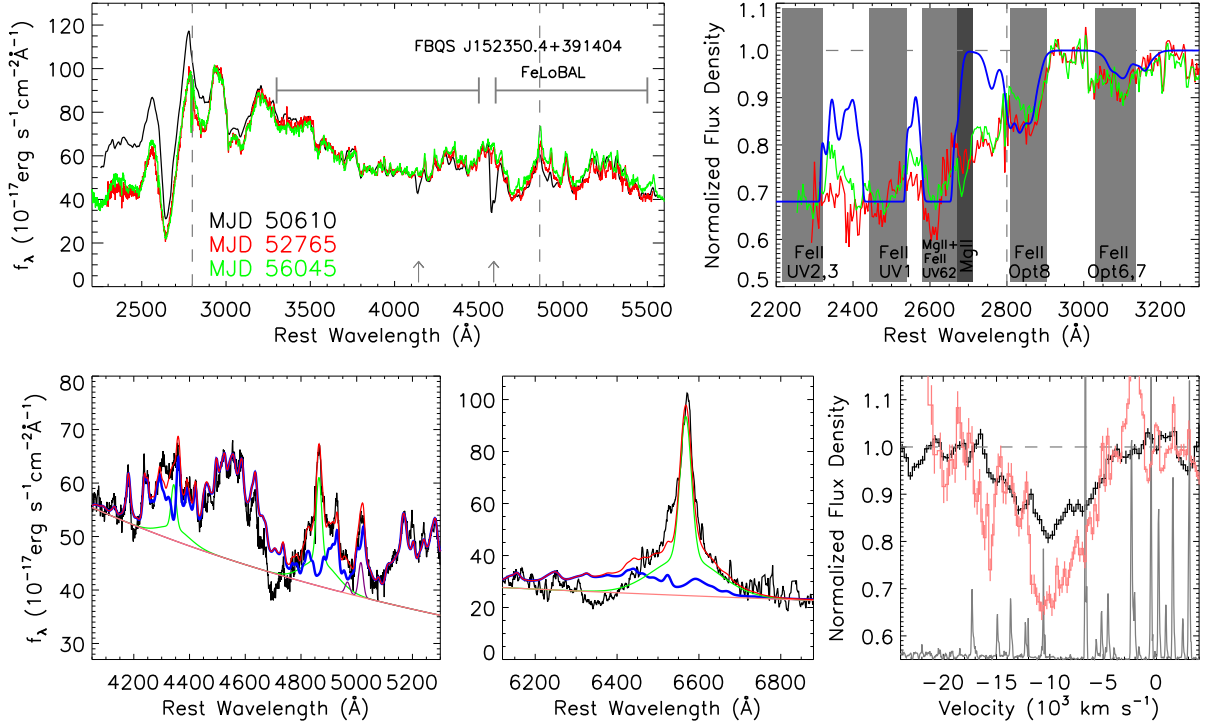}
\caption{Top panels: As Figure \ref{fig1}, but showing FBQS J152350.4+391404. 
Bottom panels: Observed spectra of H$\beta$ and H$\alpha$ regimes overplayed with the best-fit models 
and normalized spectra of Balmer absorption troughs for FBQS J152350.4+391404. 
In left- and middle-bottom panels, we plot the observed spectrum in black curves, power-law continuum in pink, 
broadened \ion{Fe}{2} template in blue, Gaussian Balmer emission lines in green, 
Gaussian \ion{O}{3} emission lines in purple, and the model sum in red. 
In right-bottom panel, the normalized spectra are obtained by dividing the observed spectra by the model spectra.
The gray dotted line means the value equals 1. Sky line spectrum (not to scale) around H$\alpha$ is plotted for comparison.
}\label{fig2}
\end{figure*}

\figurenum{3}
\begin{figure*}[tbp]
\epsscale{1.0}
\plotone{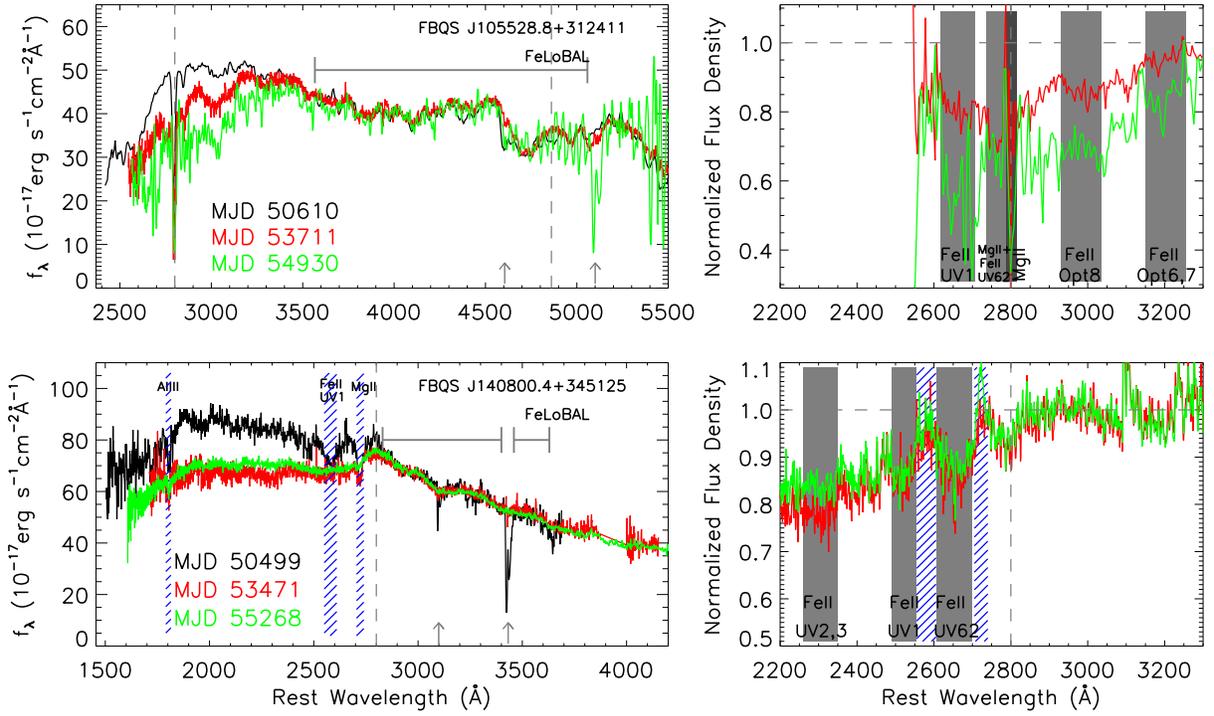}
\caption{As Figure \ref{fig1}, but showing FBQS J105528.8+12411 and J140800.4+345125. For FBQS  J140800.4+345125,
we also present the invariable broad absorption line system by blue diagonal lines.
}\label{fig3}
\end{figure*}

\figurenum{4} 
\begin{figure*}[tbp] 
\includegraphics[angle=0,width=1.0\textwidth]{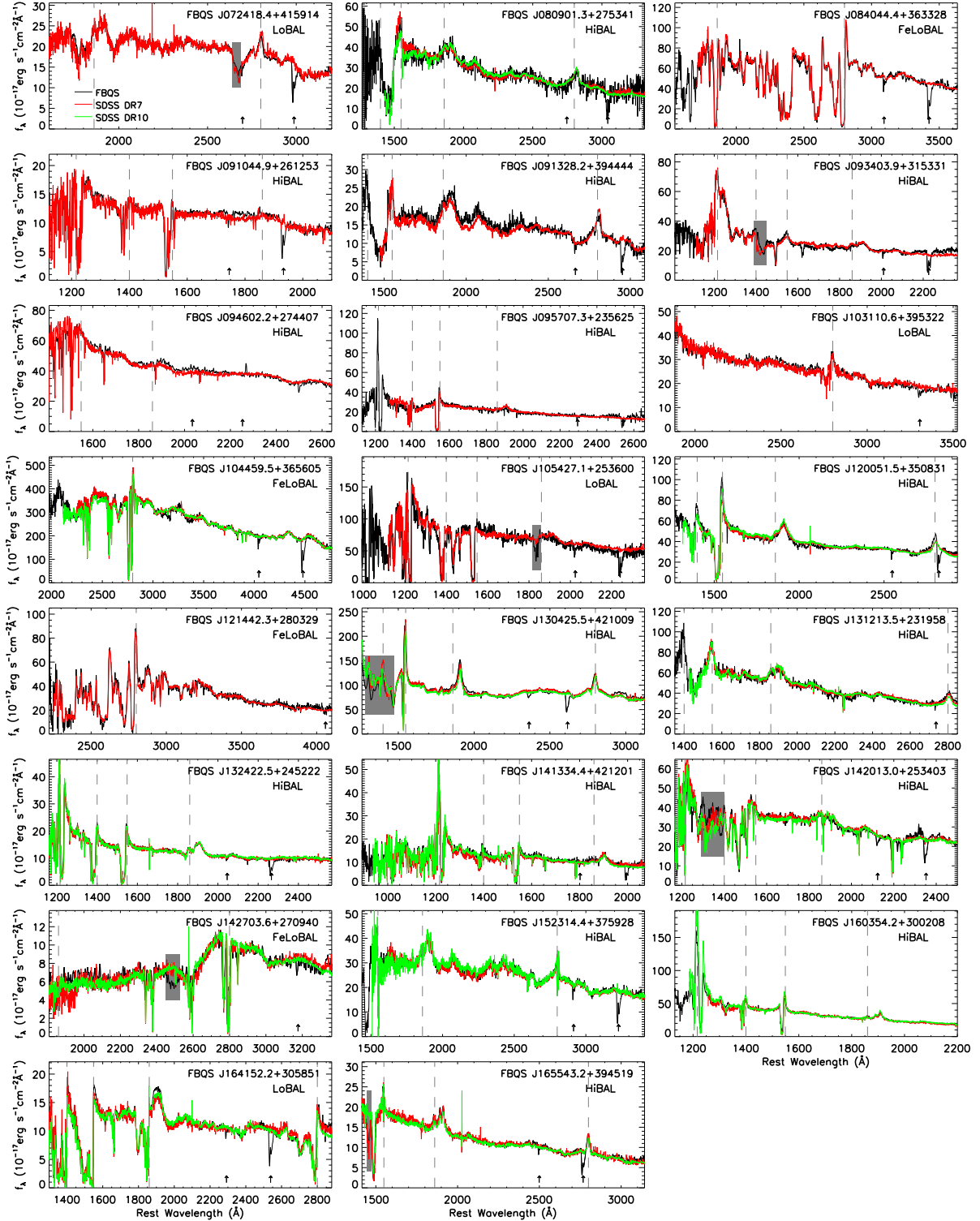}
\caption{Multi-epoch observations of 7 typical BAL variations and 16 invariable BALs in quasar's rest-frame. 
In each panel, flux scale applies to the repeated spectra to match the FBQS spectra in continuum and
spectra have been smoothed with a three-pixel boxcar.
Dashed vertical lines show the wavelengths of emission from Ly$\alpha$, \siiv, \civ, \aliii~and \mgii.
Arrows mask the potential atmospheric absorption at $\sim$6880 and 7620 \A~in the panels.
Gray shaded regions mark variable BALs for each measurement.}\label{fig4} 
\end{figure*} 

\figurenum{A1}
\begin{figure*}[tbp]
\epsscale{1.0}
\plotone{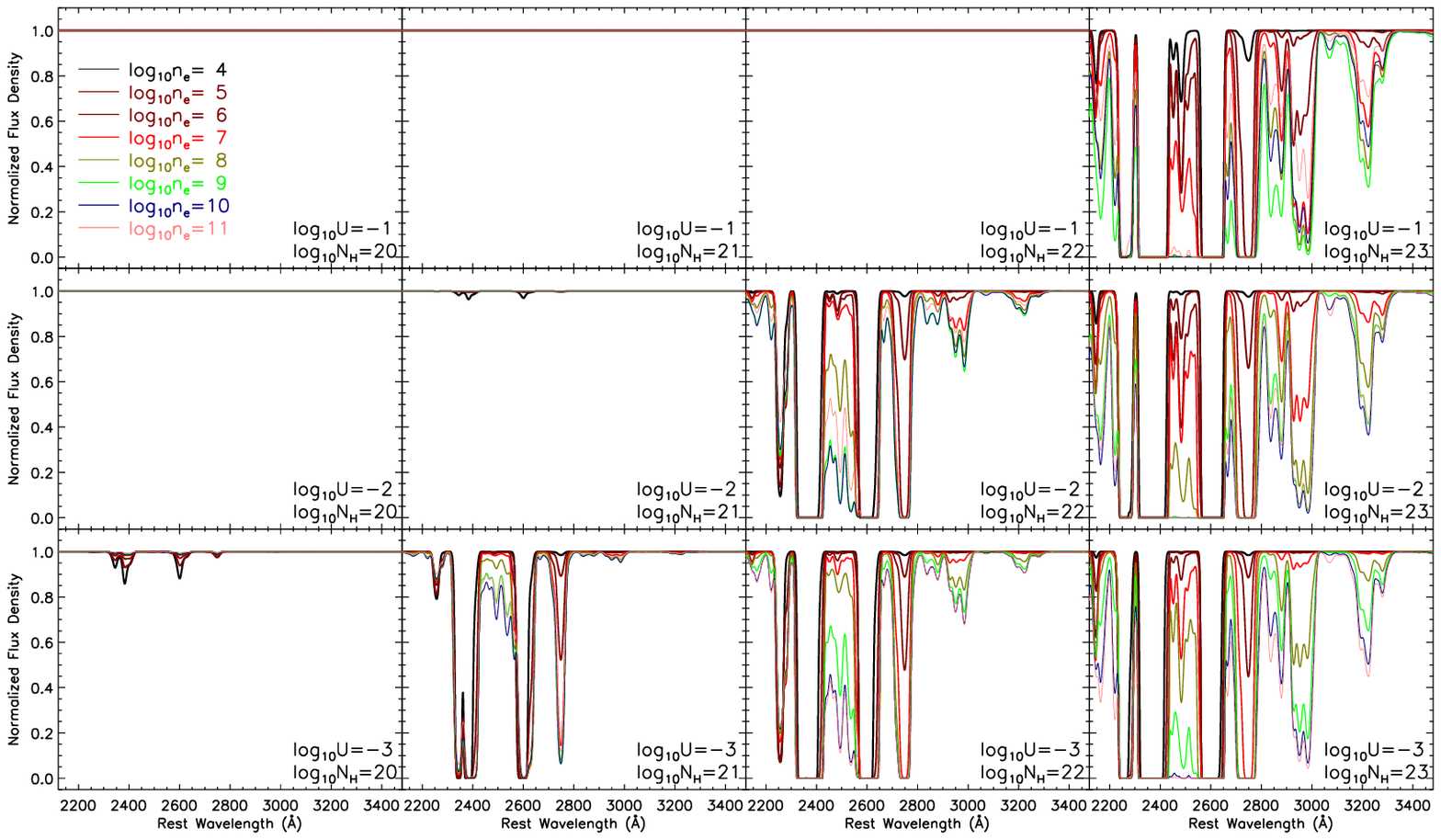}
\caption{Normalized spectra of \feii\ absorption lines in the 2100-3500 \A\ region predicted by the model using CLOUDY. 
In panels, the blueshift velocity of each \feii\ absorption line is set to 0 \kms, and their widths are 2000 \kms.
}\label{figA1}
\end{figure*}
 
\end{document}